# Archaeology of active galaxies across the electromagnetic spectrum

Raffaella Morganti

ASTRON, the Netherlands Institute for Radio Astronomy,
Postbus 2, 7990 AA, Dwingeloo, The Netherlands.
Kapteyn Astronomical Institute, University of Groningen,
P.O. Box 800, 9700 AV Groningen, The Netherlands

**Analytical and numerical galaxy formation models indicate that active galactic nuclei (AGN) likely play a prominent role in the formation and evolution of galaxies. However, quantifying their effect requires knowing how the nuclear activity proceeds throughout the life of a galaxy, whether it alternates to periods of quiescence and, if so, on which timescales these cycles occur. This topic has attracted a growing interest, but making progress remains a challenging task. For optical and radio AGN, a variety of techniques are used to perform a kind of "archaeology" to trace the signatures of past nuclear activity. Here we summarise the main recent findings regarding the AGN life cycle from optical and radio observations. The limited picture we have so far suggests that these cycles can range from long ($10^7$-$10^8$ yrs) to shorter periods of $10^4$-$10^5$ yrs, down to extreme events on timescales of only a few years. The observational results on the multiple cycles of AGN activity help, together with simulations, creating a complete picture of the AGN life cycle.**

The impact of the energy released by an AGN, i.e. an active super-massive black hole (SMBH), on the host galaxy has been recognised as a mechanism that could reconcile, throughout cosmic times, the observed number density of (massive) galaxies with that predicted by the simulations, as well as the observed relationship between black hole and bulge masses [1,2,3]. The energy produced by the AGN can remove, or at least redistribute, the gas in and around a galaxy by driving massive and fast outflows, and can prevent this gas from cooling. These processes, observed in many objects [4,5,6,7,8], can result in quenching star formation and in stopping the growth of the SMBH.

A number of optical and radio studies [9,10,11,12] have suggested that the period over which a SMBH is active is limited to a fraction of the typical lifespan of the host galaxy, i.e. the SMBH is regularly re-ignited after a period of inactivity (or low activity) and the AGN is rejuvenated. How long these periods last, is strongly dependent on the environment where the galaxy lives [13]. Defining these cycles is further complicated by the different modes of the nuclear activity [11,14,15], radiative-mode and jet-mode, fuelled by different accretion mechanisms [16] and likely characterised by different time scales. The radiative-mode drives radiatively efficient (i.e. powerful) AGN and needs dense, cold gas in the nucleus, possibly coming from a merger/interaction with other galaxies [17,18,19] or secular fuelling [14]. The jet-mode, driving radiative inefficient AGN, has been suggested to result from hot accretion or, alternatively, from accretion of cold gas clouds, as more recently proposed (chaotic cold accretion; CCA) [21,22].

However, regardless of the environment and the origin of the fuelling gas, simulations of galaxy evolution show how, even in isolated galaxies, cycles of activity are required for preventing the piling-up of gas cooling from the galaxy's atmosphere or gas resulting from stellar mass loss [23,24]. During the life of a galaxy, recurrent AGN outbursts are needed to prevent the cooling and infall of this gas which would otherwise accumulate in the central regions [21,22,32,24].

Signatures of these cycles should be present and observable in AGN, providing a record of the history of activity. What can be called *"AGN archaeology"*, i.e. trying to decipher this past record, is a challenging task, but encouraging results are appearing, in particular for optical and radio AGN. This provides exciting perspectives for the new and upcoming observing facilities.

This paper presents an overview of recent findings illustrating the variety of cycles of activity identified in radio and optical AGN: long timescales, $10^7$-$10^8$ years have been traced using radio wavelengths, and short timescales, $10^4$-$10^5$ years (and down to as short as years) at optical wavelengths. This variety is mostly the result of the different techniques used to trace and time the activity in optical and radio AGN leading to different groups of objects and different time-scales being sampled. Despite the biases, these studies are helping us in understanding how the nuclear activity is proceeding. A brief summary of some of the results from the optical studies is presented. However, most of the focus will be on radio AGN, where the new low-frequency radio telescopes, and in particular LOFAR, the LOw-Frequency Array[25] (see Box 1), is opening the possibility of major steps forward in the exploration of the life cycle of these AGN. The first results illustrating the importance of this addition and the future perspectives are described here.

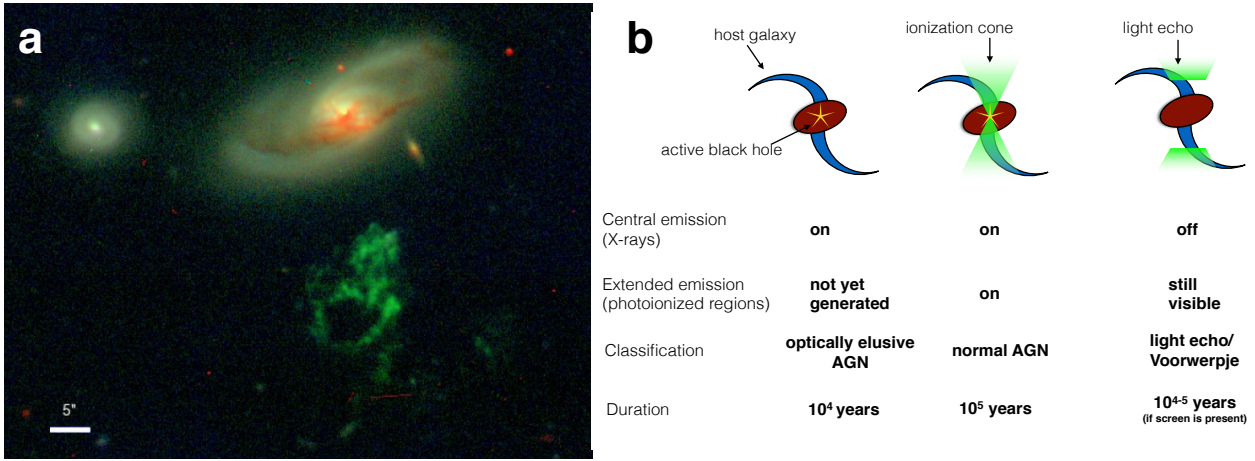

**Figure 1: The proto-typical example of "light echo" and a proposed life cycle; (a)** *Ground-based optical image of IC 2497 (top), Hanny's Voorwerp (bottom in green), and a nearby companion galaxy (left). The "green" emission of the Hanny's Voorwerp is highly ionised gas. This image is a composite of blue, visual, and near infrared light images taken with the WIYN telescope.* **(b)** *A schematic of the life cycle of optical AGN taken from (31). Panel **a** credit: WIYN / William Keel / Anna Manning. Panel **b** reproduced from ref.[31]Oxford University Press.*

## The short timescales revealed by the optical wavelengths

In optical AGN, the total length of all active phases has been derived from observations of the population of local ($z < 0.1$) SMBH. Estimating the local *relic* black-hole mass density and comparing it to the total light emitted by quasars results in a *total* accretion time ranging between $10^7$ and $10^9$ yr[10,26,27]. This constrains the total time a SMBH radiates, but does not say whether this time is made up of relatively short outbursts interleaved with periods of quiescence and, if so, how many and what are their timescales.

The discovery of clouds of highly ionised gas illuminated by an AGN near a quiescent host galaxy, and interpreted as *"light echoes"*, has made it possible to estimate the timescales of the shut-down of the optical activity. The idea of light echoes has first been proposed to explain the most spectacular case, a highly ionised region observed near IC 2497, also known as Hanny's Voorwerp[28,29]. In this galaxy (see Fig. 1 left), the lack of a nucleus bright enough to excite the ionised gas has led to the conclusion that the ionisation of the gas must be the result of a past period of nuclear activity. Given the projected separation of 20 kpc between the nucleus and the "Voorwerp", the timescale of the switching off of the AGN is estimated[28] to be ~$10^5$ yrs. Hanny's Voorwerp highlights the challenge of identifying rare objects (i.e. objects representing a phase much shorter than the typical dynamical time in a galaxy of ~Gyr) in large surveys. In the case of IC 2497 this could be done thanks to a large Citizen Science project *Galaxy Zoo*[30], aimed at classifying galaxies from the images of the Sloan Digital Sky Survey (SDSS). Galaxy Zoo and follow ups have led to the discovery of more candidates allowing to derive, based on assumptions on the occurrence and geometry of the phenomena, durations in the range $0.2$-$2\times10^5$ years for the luminous episodes of these nearby optical AGN[31].

Another indirect way of deriving the full cycle of activity has been proposed and applied to a small group of nearby AGN. The proposed method[31] (see also Fig.1 right) identifies galaxies where the SMBH has just switched on (i.e. showing an X-ray AGN but not yet an optical AGN, so called *"optically elusive AGN"*). The occurrence of these AGN suggests they switch on rapidly (~$10^4$ yr) and 'stay on' for ~$10^5$ yr before switching off[32]. As a consequence, and in order to reach the estimated total $10^7$-$10^9$ yr accretion time[10,26,27], they would have to go through a large number of these relatively short bursts, i.e. "flickering" on and *off*, with interesting implications for the accretion mechanism that would be required to fuel the black hole.

These short AGN outbursts can be modulated by longer cycles as shown by other indirect evidence. For example, the high occurrence of extended (~10 kpc) outflows found by some studies of Type 2 AGN[33] (but see also different results[34]) may suggest the presence of long episodes of activity (~$10^8$ yr) in order to inflate such a large regions. Considering it unlikely that the SMBH can keep a high luminosity for such a long time, flickering episodes (~$10^6$ yr) with intervals of ~$10^7$ yr have been suggested to describe the properties of the outflows[33].

The optical data are also telling us about the dramatic changes that an AGN can undergo on even shorter time scales, i.e. years. On such timescales, AGN are known to show various degree of variability. However, different from these are the extreme events (i.e. particularly high variation in flux) or even episodes in which they change their physical properties (e.g. with appearance or disappearance of broad Balmer lines[35, 36]). About 20 cases are known of these AGN defined as *"optical changing look"* (e.g. Mrk 1018[37]; Mrk 590[38]), now including also luminous quasars[39,40,41,42]. These objects are important for understanding the accretion processes that originates the cycle of activity of the SMBH. Most of the cases of *"optical changing look"* AGN are described as the result of a change in the *pre-existing* rate of accretion[37,40], responsible for the increase (or decrease) in the gas supply to the black hole. Tidal Disruption Events (TDE), i.e. the disruption of a star passing sufficiently close to the SMBH[43], have also been suggested as possible mechanism at the origin of changing look AGN[44]. Changing look AGN have mostly been discovered serendipitously. Only recently their search has taken advantage of dedicated spectroscopic and multi-epoch surveys, in particular SDSS and its Time Domain Spectroscopic Survey (TDSS) extension[40]. These observations have made some of these discoveries possible and are setting the requirements for future large optical surveys. The new and upcoming optical facilities and surveys that monitor the sky multiple times (like e.g. Transient Factory[45], Pan-STARRS[46] and the Large Synoptic Telescope[47]) promise a major improvement of the discovery rate and expand the statistics of changing look AGN.

The high intermittency of the activity revealed by the "flickering" and, to some respect, by the "changing look" events is also found in numerical simulations. The intermittency of the fuelling can be the result of the effects of interaction between the outflowing radiation and winds with the galactic gas, influencing and regulating the accretion[24,48]. On the other hand, in the scenario of *CCA*, the intermittency is instead connected to the discrete nature of the fuelling clouds[21,22]. According to this model, from the hot halo - ubiquitous in massive galaxies - cold clouds can condense out as result of thermal instabilities. Because of the recurrent inelastic collisions, these clouds can overcome the coherent rotation, losing angular momentum, "rain" onto the black hole and feed it. Figure 2 illustrates the variations in the accretion rate which will affect the level of activity of the SMBH[21,22]. Chaotic cold accretion can, therefore, explain the triggering of the AGN and the presence of a self-regulating loop, but also produce the rapid variations in luminosity, including the "flickering" [32] as well as *changing look* AGN. The presence of clouds of cold gas (atomic and molecular) in the regions close to the black hole has already been identified in a number of cases[49,50], supporting the presence of such a mechanism at work. Understanding the role that also TDE may have in producing the high intermittency of the activity, will require to expand the statistics and to follow the changes over longer time scales and sample a variety of time scales to overcome selection effects[43].

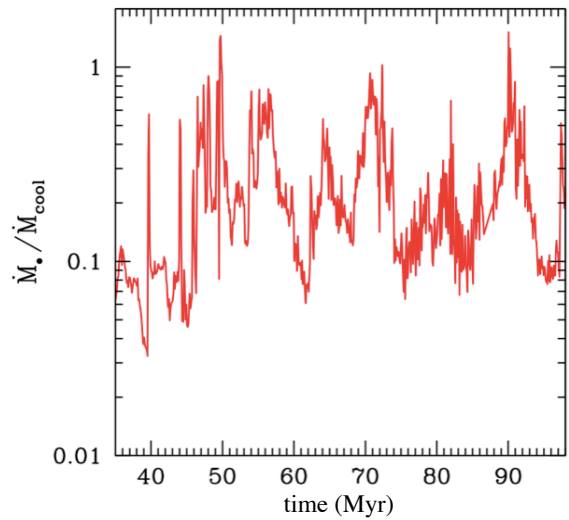

**Figure 2: The duty cycle of activity and quiescence predicted from simulations of chaotic cold accretion** *evolution of the accretion rate (including turbulence, cooling, AGN heating, and rotation) as a fraction of the cooling rate using the CCA. This illustrates the changes in the accretion rate (and therefore level of activity) on the short timescales (taken from (21), Oxford University Press.).*

# Life of a radio AGN as engraved in the radio spectrum and morphology

The fraction of galaxies having a radio-loud AGN is a strong function of the mass of the host galaxy, with up to 30% radio-loud AGN detected in massive galaxies[11]. This group of AGN has recently gained more relevance in the study of feedback because of the results of numerical simulations showing the large effect that a new or restarted radio jet can have when entering the surrounding *clumpy* ISM in the host galaxy. The expanding cocoon of shocked material created in this way, can profoundly affect a large fraction of this medium[51]. Thus, understanding how often a radio jet reignites becomes a relevant exercise.

The life of a radio-loud AGN has been explored in a number of ways, e.g. using the radio luminosity function[11,52], or the signatures in the surrounding hot medium[4,5,53] or by interpreting the features of their radio spectrum in term of ageing of the electron population. The latter benefits greatly of the advent of the new generation of low-frequency (MHz) radio telescopes like LOFAR, which are providing stronger constraints to this important region of the radio spectrum.

The emission in radio AGN is due to synchrotron radiation from relativistic electrons moving in a magnetic field[54,55]. Traditionally, this emission is considered due to a continuous injection[56] of electrons with a power-law energy distribution and resulting in a radio spectrum of the source that follows the relation $S \propto \nu^{-\alpha}$, with $S$ the observed flux and $\alpha$ the spectral index. If this is the case, for a population of electrons suffering only synchrotron losses, the ageing of the source will result in a change in the shape of the spectrum, with a frequency break appearing due to the higher radiative energy losses of higher energy electrons[57]. A second, even steeper break appears when the nuclear activity stops and the electrons are not replenished[58]. Figure 3 shows examples of how, according to this spectral ageing model, the shape of the radio spectrum can evolve over time. Following this interpretation, the modelling of the radio spectrum and its curvature (i.e. the frequencies of the breaks) can give, to first order, the active and inactive times of the source with respect to the total age. Interestingly, in this scenario, the emission at low frequencies would remain unaffected by such a changes for the longest time, acting as fossil record of the initial injection.

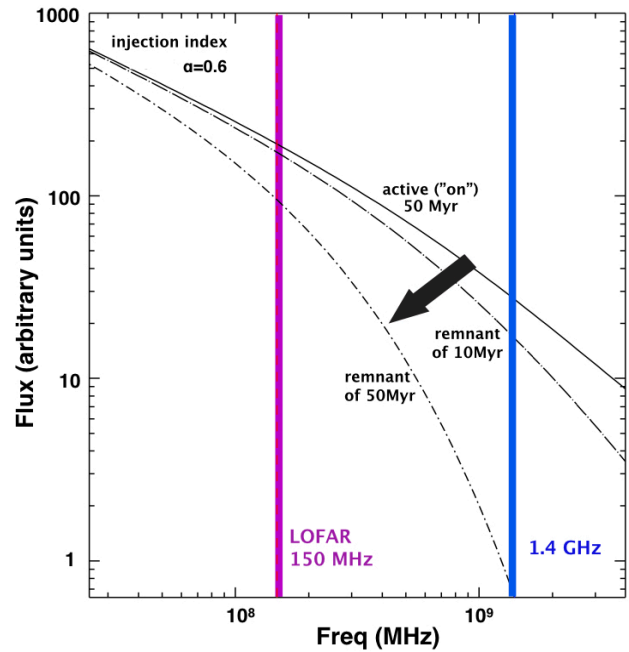

**Figure 3: Model of integrated radio spectra showing how a power-law radio spectrum is modified by the ageing of the source and by the switching off of the central AGN.** *The solid black line represents the rest-frequency spectrum of a radio source that has been "on" for 50 Myr. Because of the ageing, the electrons have suffered radiative losses, which scale with the frequency. The long dashed-dot line is the spectrum after this source has been "off" for 10 Myr while the short dashed-dot is after 50 Myr off. The arrow indicates how the spectrum changes with the increasing "off" time. Model spectra obtained using BRATS (Broadband Radio Astronomy ToolS) software[63]. The vertical lines indicate the 150 MHz frequencies covered by LOFAR and the 1.4 GHz covered by other radio telescopes.*

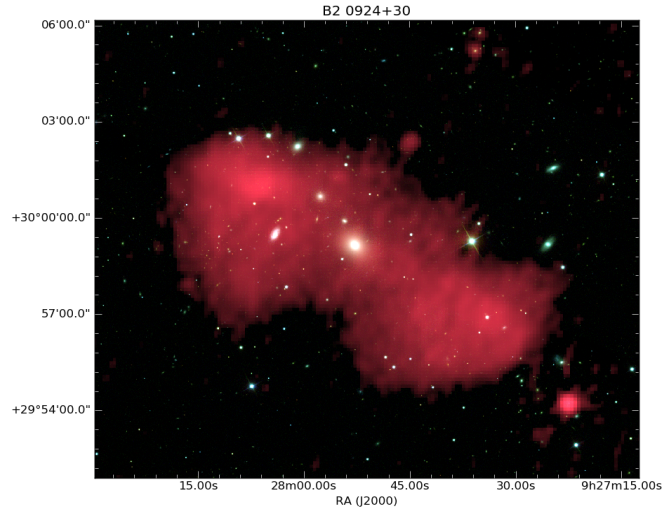

**Figure 4: An example of AGN remnant (B2 0924+30) discovered thanks to the morphology of the radio emission[82]**: *all possible signatures of on-going activity (core, jets, hot spots) are absent and only diffuse, low surface brightness emission is detected. This image of B2 0924+30 has been obtained with LOFAR at 150 MHz. In this object, the analysis of the radio spectrum[83] extended to the low frequencies reveals that the activity has ceased around 50 Myr ago. Credit: A. Shulevski.*

Given the complexity of the physics of radio sources, it is not surprising that this method is affected by large uncertainties[59,60,61,62]. They arise e.g. from the amplitude and structure (inhomogeneity) of the magnetic field[62,63], from the uncertainty in the value of the injection index of the electrons[64,65]. Differences are also found with the ages derived from dynamical arguments (which tend to be larger[59]). In addition, the frequency coverage available from the observations is typically incomplete, influencing the range of ages of the electrons that can be sampled with this method. Furthermore, alternative interpretations of the spectral steepening have also been proposed to explain the curvature observed in the radio spectra. For example, the combination of a gradient in the magnetic field and a curved initial energy spectrum of the electrons would result in a steepening consistent with what observed in the radio spectra without requiring the presence of an old electron population[59,61]. More tests are needed to investigate how well these alternative explanations can provide a general description of the observed spectra, but in the mean time this calls for some caution in the interpretation of the radio spectra.

However, so far and despite the limitations, spectral ageing has been broadly used to at least guide the interpretation of the radio spectra and the activity in radio galaxies. Follow this method, the time scales of the active phase of "adult" radio galaxies range between a few $10^7$ and $10^8$ yrs[64-68]. Their likely progenitors, young radio sources[69,70,71] with ages up to ~$10^5$-$10^6$ yrs, are identified by their small size combined with a spectrum peaked at GHz-MHz frequencies. The latter may represent the signature of synchrotron self-absorption in the phase before the source breaks away from the inner region of the host galaxy (i.e. inner few kpc). The relation between the frequency of the peak of the spectrum and the size of the source[69,72] further supports this hypothesis. Another possible interpretation of the spectral properties is free-free absorption[73,74], with the spectrum evolving as the jet-inflated bubble expands. The direct measure of the expansion of these sources[75] has shown that, at least some of, these sources will indeed evolve into large radio galaxies. However, quite a number of them will also prematurely die[76,77,78], partly as consequence of the effects of the environment known to be particularly hostile around these young sources[71,72,79]. A number of them may also represent a restarted phase of an older source as suggested by the presence of extended and older radio lobes[80,81].

*Restarted* radio sources as well as *remnants* are key objects for deriving the cycle of activity of radio galaxies. AGN remnants are the result of the switching-off of the central engine. As a consequence, all the typical signatures of activity - core, jet, hotspots - have disappeared (or are in the process of doing so), while the extended radio emission tends to become more diffuse and of low surface brightness (see Fig. 4 for an example). AGN remnants can remain *relatively* bright at low frequencies and, therefore, they are good targets to be studied with LOFAR (see below).

The best examples of sources where the nuclear activity is restarted after a period of quiescence are the so-called *double-double* radio galaxies[84]. In these objects two (or more) pairs of radio lobes, sharing the same direction of propagation, are observed. Each pair of lobes represents a phase of activity. Cases with three of such outbursts – as shown in Fig. 5 - have even been found[85]. The analysis of the radio spectra of double-doubles shows that in these objects the *off* phase tends to be comparable or shorter than the *on* phase[86,87,88], and ranges from a few to a few tens of Myr[86]. The relatively short cycle of activity is consistent with the presence, in many of them, of active hot-spots and by the restarted lobes following the same "plasma channel" of earlier bursts[87,88], perhaps a pre-requisite in order to see the source as double-double. .

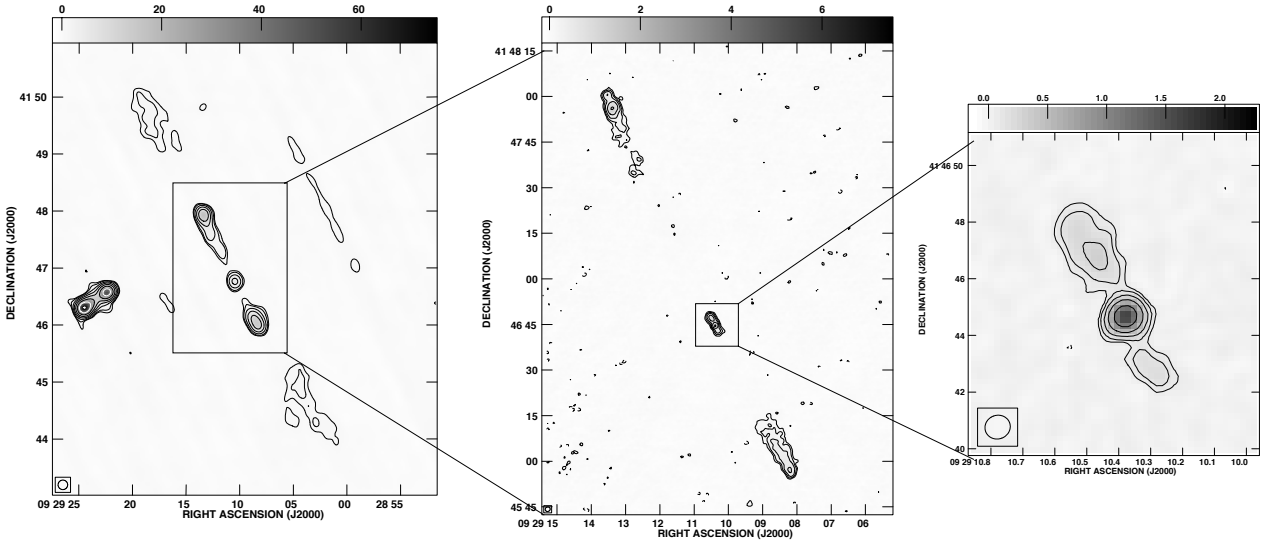

*Figure 5: A radio galaxy showing three phases of activity. The images of the radio galaxy B0925+420, obtained with the Very Large Array, are showing the three pairs of lobe[85]. In B0925+420, the age of the outer lobes was derived to be in the range 25 - 270 Myr, while that of the inner lobes is 0.4 - 2 Myr. The supply of energy for the outer lobes ceased between 4 and 70 Myr ago, while the inner lobe is still supplied by fresh electrons. The beam size is shown in the lower left corners of the plots. All contours are at −3, 3, 6, 12, 24 and 48σ, with the exception of the left-hand plot which has an additional 2σ contour. Grey-scales are in unit of mJy/beam. Oxford University Press*

# LOFAR for AGN archaeology

The revolution in the technical capabilities of low-frequency radio telescopes is providing a key addition to expand the study of remnants and restarted radio sources. The sensitive LOFAR observations in the 100 MHz band can constrain the overall spectrum of sub-mJy radio sources while providing, at the same time, high-quality, high-spatial resolution images of their morphology. This opens possibilities for the study of radio source's evolution. The technical improvements at low frequencies have started with telescopes like the Very Large Array, the Giant Metrewave Radio Telescope and, more recently, the Murchison Widefield Array[89]. These improvements have now reached new potential with LOFAR[25] (see Box 1). The capabilities of LOFAR at frequencies around 100 MHz match what has been so far routinely possible (in terms of spatial resolution and sensitivity) only for higher (GHz) frequency radio telescopes. Furthermore, the large field of view, easier to obtain at low frequencies, allows building interesting samples of rare objects.

The exploitation of the LOFAR data for the study of the life cycle of AGN has just started with the study of AGN remnants using the images and source catalogues available for famous fields, e.g. Boötes[90], H-ATLAS[91], Lockman Hole[92] combined with the available GHz data (e.g. NVSS, the NRAO VLA Sky Survey and FIRST, the Faint Images of the Radio Sky at 20 cm). A single LOFAR pointing at 150 MHz covers almost 20 sq deg of sky (see Fig. 6) and, with noise levels typically of ~0.15 mJy beam$^{-1}$, i.e. a factor 10 deeper than any previous survey at these frequencies, provides of the order of 6000 sources.

Remnants have been so far a very elusive group of objects, with not more than a few dozen of objects known over the entire sky[82,93-98]. Until now, the most common way of searching for AGN remnants was by selecting ultra-steep spectrum radio sources. As mentioned above (Fig. 3), an extremely steep spectrum (i.e. index steeper than $\alpha^{150\text{MHz}}_{1.4\text{GHz}} > 1.2$) can be used as indication of the radio emission being affected by the switching off of the central source. Thus, these sources are considered *candidate* AGN remnants. The expectation was that they would appear more numerous at low frequencies. Interestingly, despite the increased sensitivity of LOFAR, AGN remnants selected in this way remain rare, with <<6% of the radio sources belonging to this group[92,99].

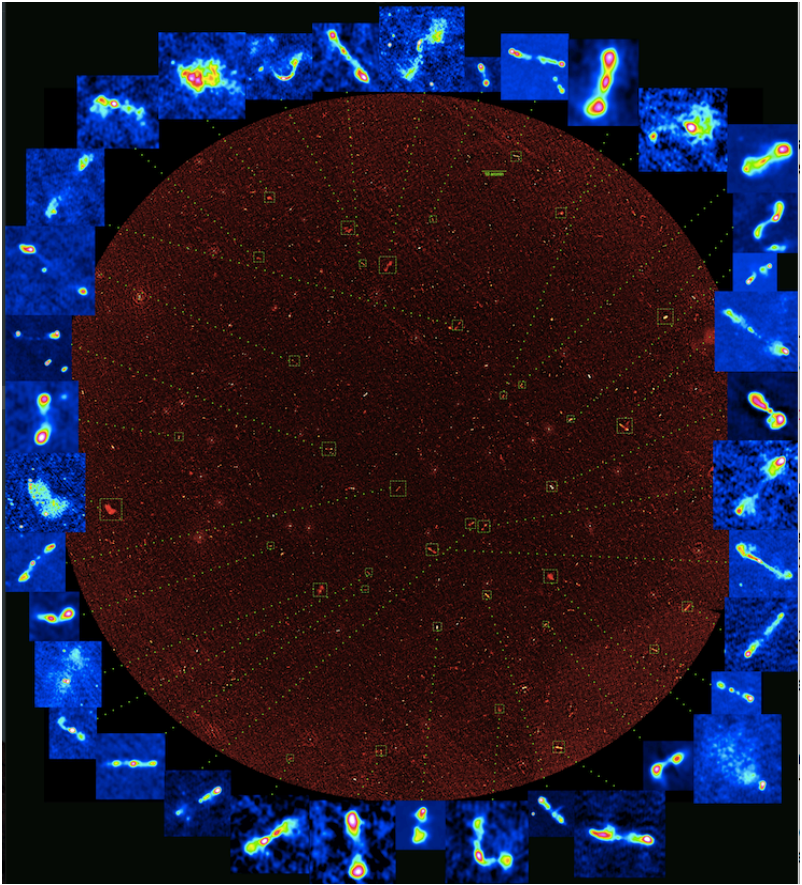

**Figure 6: LOFAR image (one pointing) centred on the Boötes area[90].** *The image at 150 MHz covers about 19 sq deg of sky with ~ 6 arcsec spatial resolution and a noise of ~ 110 microJy/beam. Zoom-ins of some representative examples of extended sources are shown to illustrate the wealth of detail on the morphology of the radio sources, including their diffuse and faint emission. Image credit: W. Williams / R. van Weeren / H. Röttgering / D. Hoang.*

Following the spectral ageing scenario, the results can be explained with a rapid luminosity evolution of the radio sources that must occur after the central engine stops, on a time scale similar to that of the steepening of the spectra. A possibility is that, when reaching the *off* phase, the radio lobes are still in over-pressure with respect to the ambient medium, causing their expansion to continue even after their active phase has ended. If, at least in some cases, this happens, the associated adiabatic losses (together with the radiative losses) could result in a fast evolution and dimming of the remnant emission. Simple modelling suggests that this could partly explain their rarity[99,100]. Thus, the remnants selected based on their steep low frequency spectrum would represent only the older remnants in the sky, with ages around $10^8$ years, while most of the remnants radio sources would be younger, i.e. in the phase shortly after the switching off of the central source (i.e. a few $\times 10^7$ yr) and with the spectrum not yet become ultra steep.

If this is the case, AGN remnants should not be such elusive objects but in order to trace them during their evolution to their final stages (~$10^8$ yr), more sophisticated selection criteria need to be used. This can now be done using the high quality of the low frequency images produced by LOFAR. Figure 6 shows an example of the wealth of detail that can be recovered for each source[85] from a LOFAR pointing centred on the Boötes area. AGN remnants are expected to be of low-surface brightness and amorphous, lacking hot-spots, jets and a central core (or with a core under-luminous compared to what observed in active sources[91,99]). Figures 4 and 7 show two examples. The quality of the LOFAR images is particularly good to allow such a morphological selection. This selection gives an higher fraction of candidate AGN remnants that, depending on the adopted criteria, can reach about 20%[91,99]. Interestingly, the majority of the morphologically selected remnants are not ultra-steep spectrum. This suggests that, unlike the ones selected based on the steep spectral index, the morphologically selected may be remnants caught soon after the switching off of their SMBH or remnants that have followed a different evolutionary path. This confirms the complementarity of the two selection methods.

The spectral properties are still needed to derive their ages and physical parameters, as done for the examples shown in Figs 4 and 7. The so-called Blob1[101] (see Fig. 7) has been discovered in one of the LOFAR fields because of its morphology and followed up at different frequencies (up to 5 GHz) and in the optical band. Its radio spectrum becomes very steep only at frequencies above 1.4 GHz. The *off* phase is estimated to be of the order of ~60 Myr while the active phase has been relatively short (15 Myr). In the case of B2 0924+30 (see Fig. 4) the analysis of the radio spectrum reveals that the activity has ceased around 50 Myr ago[83].

However, considering the limitations of the spectral ageing mentioned above, alternative scenarios should also be considered to explain the lack of ultra-steep spectrum sources. For example, it may indicate a different view of the evolution of radio galaxies, e.g. suggesting they die in a catastrophic way. If this is the case, they should have dramatic cutoff spectra, not just steep power laws, something that will need to be verified by very deep observations at higher frequencies.

While these initial results from LOFAR are not enough to time the full cycle of life of radio sources, they show that the evolution of the radio sources after their active phase can now be traced in a statistical way, key for overcoming some of the degeneracies that can be present from studies of single objects[102]. A number of steps are still required in order to obtain a complete view of the life cycle of radio AGN. For example, from the study of luminosity functions, it is expected that the duration of the remnant phase is different for high- and low-power radio sources[11,52]. Powerful radio galaxies would have longer cycles (i.e. re-triggered only once every one-to-few Gyrs), while the activity must be constantly re-triggered in lower power objects. Investigating this with the candidate AGN remnants selected from LOFAR will provide an important test for which type of radio source has likely have more impact in the feedback process[103].

Furthermore, the information derived for the AGN remnants needs to be matched with similar information (e.g. ages) for the restarted radio sources. Criteria for identifying restarted radio sources, in addition to double-double, have been proposed[96] and they now need to be applied to the LOFAR data. Building extensive statistics of AGN remnants and restarted radio sources will be ensured by the LOFAR Two-metre Sky Survey (LoTSS[104]) now in the process of imaging the entire northern sky at 150 MHz (with more than 3000 pointings). For AGN remnants, this will provide, based on the initial results, samples of *many thousands* objects.

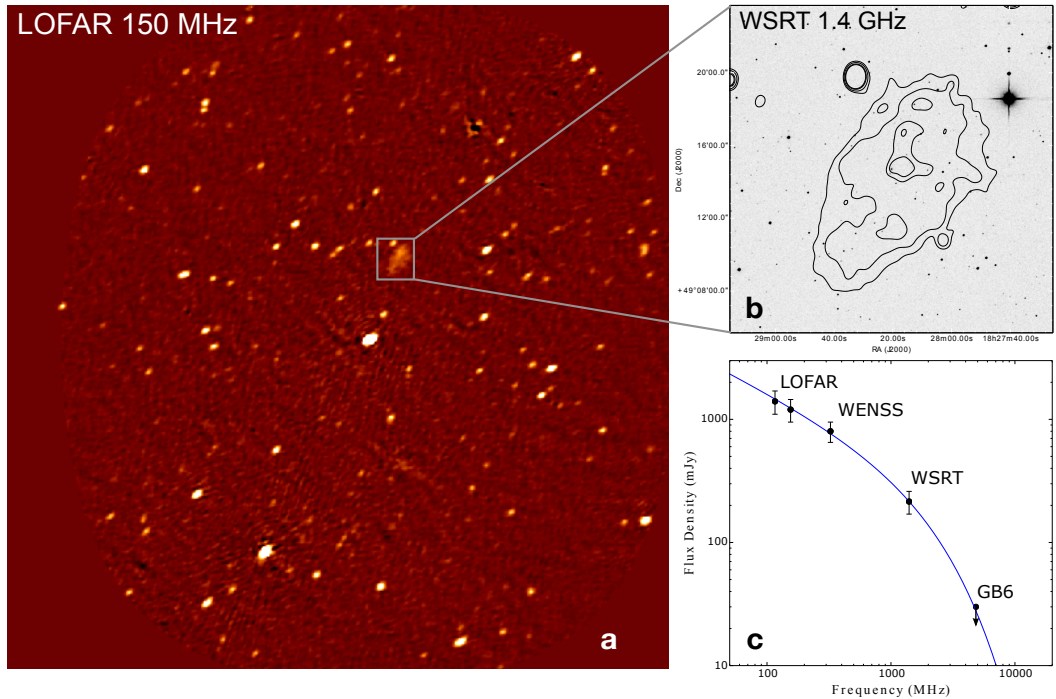

**Figure 7: A serendipitously discovered AGN remnant[101] in a LOFAR field with properties rarely seen before.**
**(a)** *The LOFAR image at 150 MHz shows the amorphous morphology of the remnant radio source.* **(b)** *This morphology is also confirmed by observations at higher frequencies.* **(c)** *The radio spectrum constructed from fluxes obtained at different frequencies with various radio telescopes (as indicated) and shown with their error bars. The best fit of the data is shown as a blue line. The radio spectrum shows a very steep slope at high frequencies indicating that the central source has turned off. However, at low frequencies the spectrum (α~0.74) is not much steeper than the original injection index. The overall shape of the spectrum is explained[101] with a source that has spent most of its life in a dying phase ($t_{off}$ = 60 Myr versus $t_{on}$ = 15 Myr). Figure adapted from ref. 101, ESO.*

# Future developments

*AGN archaeology,* using optical and radio observations, is providing constraints on the cycle of nuclear activity and on the models of AGN accretion and feedback. Numerical simulations are starting to describe the observed characteristics of the life cycle of AGN and to provide the tools needed for a more quantitative comparison with the data, something to look forward to in the near future. However, a comprehensive picture of these phenomena requires a systematic search for AGN in different phases of their evolution to expand the small samples, mostly limited to serendipitous discoveries, available so far. The new and upcoming large surveys will ensure, together with techniques for selecting rare objects or large Citizen Science Projects like Galaxy Zoo and Radio Galaxy Zoo[105], a major expansion in the discovery rate and the next big steps forward in this field. The role of surveys which revisit the same part of the sky with high cadence is key for expanding the number of known "*optical changing look*" AGN. The improved survey speed of the new radio telescopes will allow to explore the *flickering* and *changing look* phenomena also at radio wavelengths where only very limited information is available on this short time-scale events.

The good image quality of the low-frequency surveys is opening up exciting possibilities for exploring in depth the evolution of radio sources by combining the morphology of the radio emission with the spectral properties for large sample of objects in different phases of their evolution. Spatially resolved studies, with good multi-frequency coverage, will allow to trace the electrons populations of different ages and their energetics. The spectral properties will become even more accurately known with the addition of new, deep GHz surveys about to become available from the new generation of radio telescopes (i.e. the SKA pathfinders). Last but not least, the synergy between the large optical and high-resolution, radio surveys is the next challenge to face. Being able to obtain optical identifications of radio sources and quantify their distribution in redshift and radio power is a key step to unravel their physical properties and their evolution.

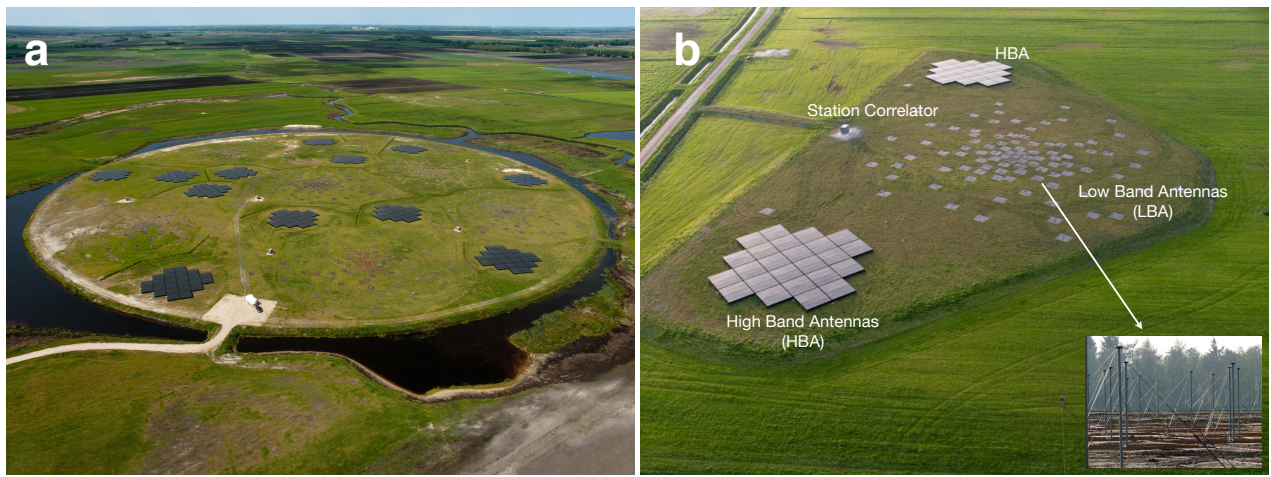

**Figure 8: Two views of the LOFAR radio telescope; (a)** *LOFAR "superterp", an area of 300 m hosting 6 stations and located in the north of the Netherlands;* **(b)** *One of the 38 Dutch stations of LOFAR. The two types of antenna are marked: HBA operating around 150 MHz and LBA operating around 60 MHz (with a zoom-in of the antenna in the inset). The small grey box represents the station correlator where the beam forming and the first correlation of the signal from the dipoles in the field is happening. Image credit: © Top-Foto, Assen.*

## Box 1 | LOFAR, the Low Frequency ARray

Although radio astronomy started with low frequency (MHz) instruments (with the work of Grote Reber) because of the simplicity of the receivers, it is only in recent years that the technical improvements in computing capabilities have made it possible to construct a new generation of low frequency radio telescopes reaching high spatial resolution and high sensitivity. These instruments (e.g. GMRT, VLA, LWA, MWA and LOFAR) are now paving the way to the low-frequency part of the Square Kilometre Array. Of the available low frequency instruments, LOFAR, the LOw-Frequency Array[25], is the one reaching the highest sensitivity and the highest spatial resolution. LOFAR is a radio interferometer which consists of dipole antenna stations distributed throughout the Netherlands and extending further into Europe. It operates in the frequency range 10 to 240 MHz (corresponding to wavelengths of 30 to 1.2 m) using two types of antennas: low-band, LBA (operating around 60 MHz) and high-band, HBA (operating around 150 MHz). Figure 8 right shows one of the 38 Dutch stations including both HBA and LBA antennas. The array using the HBA has a collecting area of almost 75000 m$^2$ with 47616 dipoles pairs.

LOFAR can be defined as a "software telescope" because it has no movable parts and the pointing and beam forming are done in software. Unique to LOFAR is the spatial resolution, which is unprecedented for a low frequency telescope, combined with the huge Field-of-View (many sq deg). Spreading out from a central "core" in the northeast of the Netherlands (Fig. 8 left), it includes baselines up to about 80 km, reaching spatial resolutions of ~5 arcsec at 150 MHz. The international stations distributed in different countries in Europe (i.e. Germany, UK, France, Sweden, Poland and soon Ireland) allow to reach sub-arcsec resolution. The strength of LOFAR is in the computing capabilities: the data flow from all antennas combined is 1.7 TByte/s which is reduced a factor more than 1000 in real time by the beam former and correlation. One of the main projects carried out by LOFAR is the statistical detection of the HI signal of the Epoch of Reonization. Among the other large projects are the deep continuum surveys of the northern hemisphere and the study of the transient sky. In addition to these, LOFAR also operates as open-time telescope.

**Acknowledgements**

*I would like to thank Tom Oosterloo, Massimo Gaspari, Leith Godfrey, Marisa Brienza, Jeremy Harwood, Aleksandar Shulevski, Luca Ciotti and the LOFAR Survey Team for help, comments and discussions. The research leading to these results has received funding from the European Research Council under the European Union's Seventh Framework Programme (FP/2007-2013) / ERC Advanced Grant RADIOLIFE-320745. LOFAR, the Low Frequency Array designed and constructed by ASTRON, has facilities in several countries, that are owned by various parties (each with their own funding sources), and that are collectively operated by the International LOFAR Telescope (ILT) foundation under a joint scientific policy.*

**Autor information**
Raffaella Morganti    *morganti@astron.nl*
ASTRON, the Netherlands Institute for Radio Astronomy,
    Postbus 2, 7990 AA, Dwingeloo, The Netherlands.
Kapteyn Astronomical Institute,  University of Groningen,
    P.O. Box 800, 9700 AV Groningen, The Netherlands

**Correspondence and requests for materials should be addressed to RM**

**Competing interests**

 The authors declare no competing financial interests.